\documentclass{desyproc}

\def\un#1{\,{\rm #1}}
\def\unt#1{\,{\rm (#1)}}
\def\der{{\rm d}}

\begin{document}
\title{TOTEM Results on Elastic Scattering and Total Cross-Section}

\author{{\slshape Jan Ka\v spar} of behalf of the TOTEM collaboration\\[1ex]
CERN, 1211 Geneva 23, Switzerland,\\
Institute of Physics, AS CR, v.v.i., 182 21 Prague 8, Czech Republic
}

 
\acronym{EDS'13} 

\maketitle

\begin{abstract}
TOTEM is an LHC experiment dedicated to forward hadronic physics. In this contribution, two main parts of its physics programme -- proton-proton elastic scattering and total cross-section -- are discussed. The analysis procedures are outlined and their status is reviewed.
\end{abstract}

\section{The TOTEM experiment}

The TOTEM experiment \cite{totem_jinst} is dedicated to forward hadronic phenomena at the LHC. The three pillars of its physics programme are: an accurate determination of the total cross-section, a measurement of elastic scattering in a wide kinematic range and studies of diffractive processes. This paper is focused on the first two. 

The physics programme brings special requirements for the detector apparatus. In particular, large pseudorapidity coverage (to detect most fragments from inelastic collisions) and excellent acceptance for outgoing diffractive and elastic protons. To accomplish this task, TOTEM comprises three subdetectors: the inelastic telescopes T1 and T2 and a system of Roman Pots (RP) for leading proton detection.

The RP system consists of two stations placed at $+220\un{m}$ (right arm) and $-220\un{m}$ (left arm) from the LHC interaction point (IP) 5. Each of the stations is composed of two units (far and near with respect to the IP) separated by about $5\un{m}$, which is beneficial for reconstructing proton kinematics and discrimination from background. Each unit includes two vertical (top and bottom) and one horizontal RP. The RPs are movable beam-pipe insertions that can bring sensitive detectors to sub-millimetre distance from the beam once it is stable. Each RP hosts 5 back-to-back mounted pairs of silicon strip sensors with reduced ($\approx 50\un{\mu m}$) insensitive margin on the edge facing the beam, in order not to loose protons scattered to very low angles.

\section{Elastic scattering}\label{s:el}

\subsection{Analysis steps}

The elastic scattering final state consists of two (almost) anti-collinear protons that are detected in the vertical RPs. Therefore the recorded events split into two ``diagonal'' topologies. In one topology, the proton in the left arm hits the bottom RP and the proton in the right arm the top RP, and vice versa in the other topology. As there are different detectors involved in the two diagonals, they can be regarded as different experiments (although not fully independent as they share the same beams). Furthermore, a usual elastic scattering analysis is performed with data from several LHC fills and/or RP position, which increases the control over possible systematic effects.

In the paragraphs below, a typical analysis will be outlined. Here, we would like to anticipate one of its unique features -- it is purely data-driven. No Monte-Carlo with a physics model is exploited in any of the steps.

{\em Alignment}. Three complementary methods are generally applied. Before data taking, the RP positions are calibrated by the beam-based alignment -- similar procedure as for LHC collimators. The off-line alignment proceeds in two steps. First, proton tracks passing through the overlap between the vertical and horizontal RPs are used to determine the relative alignment of the RPs of each unit. Second, since the elastic event tagging (see below) does not require a precise alignment, an alignment fine-tuning is performed on a sample obtained from a pre-selection of elastic events. By exploiting the azimuthal symmetry of elastic scattering, the horizontal and vertical shifts and the tilt of each unit are adjusted.

{\em Kinematics reconstruction}. Proton scattering angles and vertex positions are reconstructed from the hits detected in RPs using LHC optics. Good knowledge of the optics is, therefore, critical. Here it should be noted that TOTEM has developed a technique to fine-correct the optical functions based on the RP observables \cite{optics}. As anticipated above, the two RP units per station are advantageous since the reconstruction formulae can be optimised in order to minimise systematic uncertainties. For instance, the vertical scattering angle is typically reconstructed directly from the hit position while the horizontal angle from the local track angle at the RP station.

{\em Elastic tagging}. The basic selection cuts enforce the elastic event topology: two anti-collinear protons (required in both transverse projections $x$ and $y$) coming from a common vertex (usually only the $x$ component can be reconstructed). Furthermore, cuts to require elasticity (low fractional momentum loss, $\xi$) can be applied, for example by exploiting the position-angle correlation (commonly only in $y$ direction).

{\em Background subtraction}. The above elastic tagging has normally very good efficiency (negligible amount of true elastic events lost) but possibly non-perfect purity (non-elastic events passing the selection, i.e.~background). Often, the background is very low for high $\beta^*$ optics since the beam divergence is rather small and therefore the collinearity cuts (see above) are very powerful. In general, the background estimation method is based on interpolating the event distribution from the region surrounding the signal (tagged) region.

{\em Acceptance corrections}. Typically, there are two acceptance limitations identified in the data: the detector edge (relevant for low $|t|$) and the LHC aperture limitations (relevant for high $|t|$). Both effects are treated by assuming azimuthal symmetry of the elastic scattering (verified in the portion of data directly accessible) leading to a purely geometrical correction. In addition, there is a correction for smearing around the limitation edges.

{\em Unfolding of resolution effects}. The angular resolution is set by beam divergence (thus better for high $\beta^*$ optics) and for the $x$ projection also by the resolution of the RP silicon sensors. The angular resolution can be determined experimentally by comparing the proton scattering angles reconstructed from the left and right arm. The resolution impact on the differential cross-section is determined and eliminated by an iterative procedure that starts by taking the observed (smeared) $t$-distribution as an input to a Monte-Carlo calculation of a\break per\discretionary{-}{-}{-}bin un-smearing correction. The correction is applied to the observed $t$-distribution and yields a better estimate of the true $t$-distribution. These two steps are repeated usually three times to reach convergence.

\eject

{\em Inefficiency corrections}. The RP trigger efficiency is controlled with the zero-bias data stream -- the trigger logic is applied a posteriori in software and compared to the trigger decision flag stored in data. Generally, the inefficiency is negligible. The DAQ inefficiency (dead time) is determined by comparing the numbers of triggered and recorded events, typically yielding corrections on a percent level. Reconstruction inefficiencies may occur due to: intrinsic RP detection inefficiency of each silicon sensor, proton interactions with the material of a RP and ``pile-up'' of several particles in one event (RPs can uniquely reconstruct one track only). For the latter case, the most important contribution is a coincidence of an elastic and a beam-halo proton.

Uncorrelated inefficiencies of individual RPs are studied by removing the RP in question from the selection cuts and counting the recovered events. For this study, only a subset of the selection cuts is retained (some cuts require both near and far RP measurements). A typical result for high $\beta^*$ optics is an inefficiency about $1.5\%$ ($3\%$) for near (far) RPs. This difference can be explained by proton interactions in the near RP that affect the far RP too. This near-far correlated inefficiency is determined from data by counting events with corresponding shower signatures, yielding about $1.5\%$, which is confirmed by Monte-Carlo simulations.

The ``pile-up'' inefficiency is calculated from the probability of finding an additional track (on top of the elastic) in any station of a diagonal. This probability is determined from the zero-bias data stream and was often found to increase with closer RP approach to the beam or with degrading beam conditions. The pile-up probabilities may reach up to about $10\%$.

{\em Luminosity}. For the $\sqrt s = 7\un{TeV}$ analyses (see below), the luminosity was provided by CMS (with a $4\%$ uncertainty). For the $\sqrt s = 8\un{TeV}$ analyses, the luminosity was determined by TOTEM (also with a $4\%$ uncertainty) as a by-product of the total cross-section measurement by the luminosity-independent method, see Eq.~(\ref{eq:si tot m3}).

{\em Study of systematic uncertainties}. For each of the analysis steps above, the systematic uncertainty effect
on differential cross-section is estimated with a Monte-Carlo simulation, using the final experimental $t$-distribution as an input. Quite generally, the luminosity uncertainty is the leading $t$-independent contribution. Regarding the $t$-dependent ones, the uncertainty due to residual misalignment tends to dominate at low $|t|$ while the optics uncertainties (affecting the kinematics reconstruction) at higher $|t|$.

\subsection{Overview of analyses and results}

There are published or ongoing elastic scattering analyses at centre-of-mass energies $\sqrt s = 7$, $8$ and $2.76\un{TeV}$. For details on the used data samples, $|t|$ ranges, event statistics and analysis status/publication reference see Tables \ref{tab:el res 7}, \ref{tab:el res 8} and \ref{tab:el res 2.76}.

\begin{table}
\caption{List of elastic scattering analyses at $\sqrt s = 7\un{TeV}$. The LHC optics is characterised by the betatron function value at the IP, $\beta^*$. The RP approach to the beam is given in multiples of the transverse beam size, $\sigma$. The number of elastic events corresponds to both diagonals after the tagging.
}
\label{tab:el res 7}
\begin{center}
\begin{tabular}{|c|c|c|c|c|}\hline
$\beta^*\unt{m}$ & RP approach & $|t|$ range$\unt{GeV^2}$ & elastic events & status\cr\hline
\hline
$90$  & $10\un{\sigma}$					& $0.02\hbox{ to } 0.4$ 		& $15\un{k}$	& published \cite{si_el_7_90a}\cr\hline
$90$  & $4.8 \hbox{ to }6.5\un{\sigma}$	& $0.005\hbox{ to } 0.4$ 		& $1\un{M}$		& published \cite{si_el_7_90b}\cr\hline
$3.5$ & $7\un{\sigma}$					& $0.4\hbox{ to } 2.5$			& $66\un{k}$	& published \cite{si_el_7_3p5}\cr\hline
$3.5$ & $18\un{\sigma}$					& $\approx 2\hbox{ to } 3.5$	& $10\un{k}$	& analysis advanced\cr\hline
\end{tabular}
\end{center}
\end{table}

\begin{table}
\caption{List of elastic scattering analyses at $\sqrt s = 8\un{TeV}$, the same legend as in Table \ref{tab:el res 7}.}
\label{tab:el res 8}
\begin{center}
\begin{tabular}{|c|c|c|c|c|}\hline
$\beta^*\unt{m}$ & RP approach & $|t|$ range$\unt{GeV^2}$ & elastic events & status\cr\hline
\hline
$1000$	& $3\hbox{ or }10\un{\sigma}$	& $0.0006\hbox{ to }0.2$	& $352\un{k}$	& publication in preparation\cr\hline
$90$	& $6\hbox{ to }9.5\un{\sigma}$	& $0.01\hbox{ to }0.3$		& $0.68\un{M}$	& completed (cf.~\cite{si_tot_8})\cr\hline
$90$	& $9.5\un{\sigma}$				& $0.02\hbox{ to }1.4$		& $7.2\un{M}$	& analysis advanced\cr\hline
\end{tabular}
\end{center}
\end{table}

\begin{table}
\caption{List of elastic scattering analyses at $\sqrt s = 2.76\un{TeV}$, the same legend as in Table \ref{tab:el res 7}. The $|t|$ range is, so far, indicative only.}
\label{tab:el res 2.76}
\begin{center}
\begin{tabular}{|c|c|c|c|c|}\hline
$\beta^*\unt{m}$ & RP approach & $|t|$ range$\unt{GeV^2}$ & elastic events & status\cr\hline
\hline
$11$ & $5\un{\sigma}$	& $0.06\hbox{ to }0.4$	& $45\un{k}$	& analysis in progress\cr\hline
$11$ & $13\un{\sigma}$	& $0.4\hbox{ to }0.5$	& $2\un{k}$		& analysis in progress\cr\hline
\end{tabular}
\end{center}
\end{table}

\subsection{Studies of Coulomb-nuclear interference}

The $8\un{TeV}$ data taken at $\beta^* = 1000\un{m}$ optics permit to determine the elastic differential cross-section down to $|t| = 6\cdot10^{-4}\un{GeV^2}$. Thanks to that, the interference between electromagnetic (traditionally called Coulomb) and strong (called nuclear) interactions has been observed for the first time at the LHC. Studying the interference is interesting for (at least) two reasons: it gives sensitivity to the phase of the nuclear amplitude (only possibility of measurement beyond the cross-section level) and allows to separate the Coulomb and nuclear effects (beneficial for determinations of the {\em nuclear} total cross-section, see Section~\ref{s:tot}).

In theory, the interference is described by an interference formula combining the Coulomb scattering amplitude ${\cal A}^{\rm C}$ and nuclear amplitude ${\cal A}^{\rm N}$ into a complete amplitude ${\cal A}^{\rm C+N}$ that accounts for both interactions acting simultaneously:
\begin{equation}\label{eq:int f}
{\cal A}^{\rm C+N} = \hbox{interference formula}({\cal A}^{\rm C}, {\cal A}^{\rm N})\ .
\end{equation}
Historically, there have been two approaches to derive the interference formula: pursuing Feynman diagram technique (best represented by the simplified formula by West and Yennie \cite{wy68}, here abbreviated SWY) and using eikonal description (most recently represented by Kundr\' at and Lokaj\' i\v cek \cite{kl94}, denoted KL). The SWY formula is only consistent with purely exponential nuclear amplitudes with constant phase. No such explicit limitations apply to the KL formula.

While the Coulomb amplitude ${\cal A}^{\rm C}$ is well known (e.g.~from QED with experimental form factors), the nuclear amplitude ${\cal A}^{\rm N}$ cannot be calculated from first principles. Still, its modulus $|{\cal A}^{\rm N}|$ is significantly constrained by the TOTEM differential cross-section measurements. Therefore, in TOTEM analyses, it is parametrised by an exponential form $a\,\exp(b_1 t + b_2t + \ldots)$ with the number of $b_i$ parameters $N_b$ between 1 and 3. There is no similar constraint for the phase of the nuclear amplitude $\arg {\cal A}^{\rm N}$, thus several theoretical alternatives are considered: constant, central and peripheral phase. The latter two are motivated by different behaviours in the impact parameter space \cite{kl94}. In the central case, elastic collisions are more central than inelastic ones, and vice versa for the peripheral.

TOTEM has adopted an explorative approach to the analysis and treat all the above mentioned options on equal footing. The analysis is performed by generalised $\chi^2$ fits (accounting for full statistical and systematic uncertainty bands) through the entire $|t|$ range. Typical $\chi^2/\hbox{ndf}$ is very close to $1$. Regarding the phase of the nuclear amplitude, only its value at $t = 0$ is kept free in the fits. This value is related to the rho parameter $\rho \equiv \Re {\cal A}^{\rm N} / \Im {\cal A}^{\rm N}$.

Although the analysis is not yet completed, the preliminary results show quite convincingly that the fits can not distinguish between constant and central phase. Similarly, no significant difference has been observed between the fits with $N_b = 1$ and SWY or KL interference formula. The fit results for $\sigma_{\rm tot}$ are very stable (little dependence on the chosen description options) and are about $1\un{mb}$ higher than the previous result \cite{si_tot_8}, therefore very well compatible.

\section{Total cross-section}\label{s:tot}

\subsection{Determination methods}

TOTEM has applied three different methods to determine the total cross-section.

\begin{itemize}

\item The first method exploits only elastic inputs (thus measured by RPs). By applying the optical theorem, once can obtain the following formula for the total cross-section $\sigma_{\rm tot}$:

\begin{equation}\label{eq:si tot m1}
\sigma_{\rm tot}^2 = {16\pi\over 1+\rho^2} {1\over {\cal L}} \left. {\der N_{\rm el}\over \der t}\right |_{t = 0}\ ,
\end{equation}
where ${\cal L}$ stands for the integrated luminosity and $\der N_{\rm el}/\der t|_{t = 0}$ is the elastic differential rate extrapolated to $t = 0$.

\eject

\item The second method relies on summing the elastic event rate $N_{\rm el}$ (obtained by integrating and extrapolating the differential rate) and inelastic event rate $N_{\rm inel}$ (measured by the inelastic telescopes T1 and T2, see below):
\begin{equation}\label{eq:si tot m2}
\sigma_{\rm tot} = {1\over {\cal L}} (N_{\rm el} + N_{\rm inel})\ .
\end{equation}
This method does not require the value of $\rho$ which is advantageous when TOTEM cannot determine $\rho$ itself and needs to take it from an external source.

\item The third method is called luminosity-independent as the inputs are combined in such a way that the formula for $\sigma_{\rm tot}$ does not include the luminosity. Moreover, the value of the integrated luminosity can be determined as well:

\begin{equation}\label{eq:si tot m3}
\sigma_{\rm tot} = {16\pi\over 1+\rho^2} {\der N_{\rm el}/ \der t |_{t = 0}\over N_{\rm el} + N_{\rm inel} }\ ,\qquad
{\cal L} = {1+\rho^2\over 16\pi} {( N_{\rm el} + N_{\rm inel})^2 \over \der N_{\rm el}/ \der t |_{t = 0}}\ .
\end{equation}

\end{itemize}

The principle of the elastic scattering analysis is explained in Section \ref{s:el}. Here, we would like to outline the measurement of the inelastic event rate $N_{\rm inel}$, for more details we refer the reader elsewhere \cite{si_tot_8,si_inel_7}. The measurement proceeds in three steps. In the first step, {\em raw rate} is established by counting events with the telescope T2 (fragments of more than $95\un{\%}$ of inelastic events are detected). In the second step, one applies corrections to determine {\em visible rate} that would be measured with perfect detector in perfect conditions. These corrections account for trigger and reconstruction inefficiencies, beam-gas background and event pile-up. In the third step, events with no tracks in T2 are recovered:  T1-only events, events with rapidity gap over T2, low-mass diffraction and central diffraction without tracks in T1 and T2. The only major Monte-Carlo based correction in the analysis is the one for low-mass diffraction. However, it should be noted that TOTEM can constrain its value from data by calculating $\sigma_{\rm tot}^{\rm RP} - \sigma_{\rm el}^{\rm RP} - \sigma_{\rm inel, visible}^{\rm T2}$, where the first two terms (the last one) are determined with RPs (T2).

\subsection{Overview of analyses and results}

At $\sqrt s =7\un{TeV}$, two total-cross section analyses were published, both exploiting the $\beta^*=90\un{m}$ optics. The first paper \cite{si_el_7_90a} (cf.~top row in Table~\ref{tab:el res 7}) used the method based on elastic inputs only, Eq.~(\ref{eq:si tot m1}). The second publication \cite{si_tot_7} used a dataset (second row from top in Table~\ref{tab:el res 7}) with much higher statistics and all the three methods were applied. Let us remark that all four total cross-section results are well compatible.

At the energy of $8\un{TeV}$, the luminosity-independent results on elastic, inelastic and total cross-section were published \cite{si_tot_8}. Moreover, the analysis of the $\beta^*=1000\un{m}$ data is in progress. With these data, the separation of Coulomb and nuclear effects is at reach, thus yielding methodically more accurate results.

At $\sqrt s = 2.76\un{TeV}$ TOTEM aims at applying all three total cross-section determination methods. The inelastic part of the analysis is almost completed, the elastic part is ongoing.

\begin{figure}
\begin{center}
\leavevmode
\hbox{\pdfximage width15cm {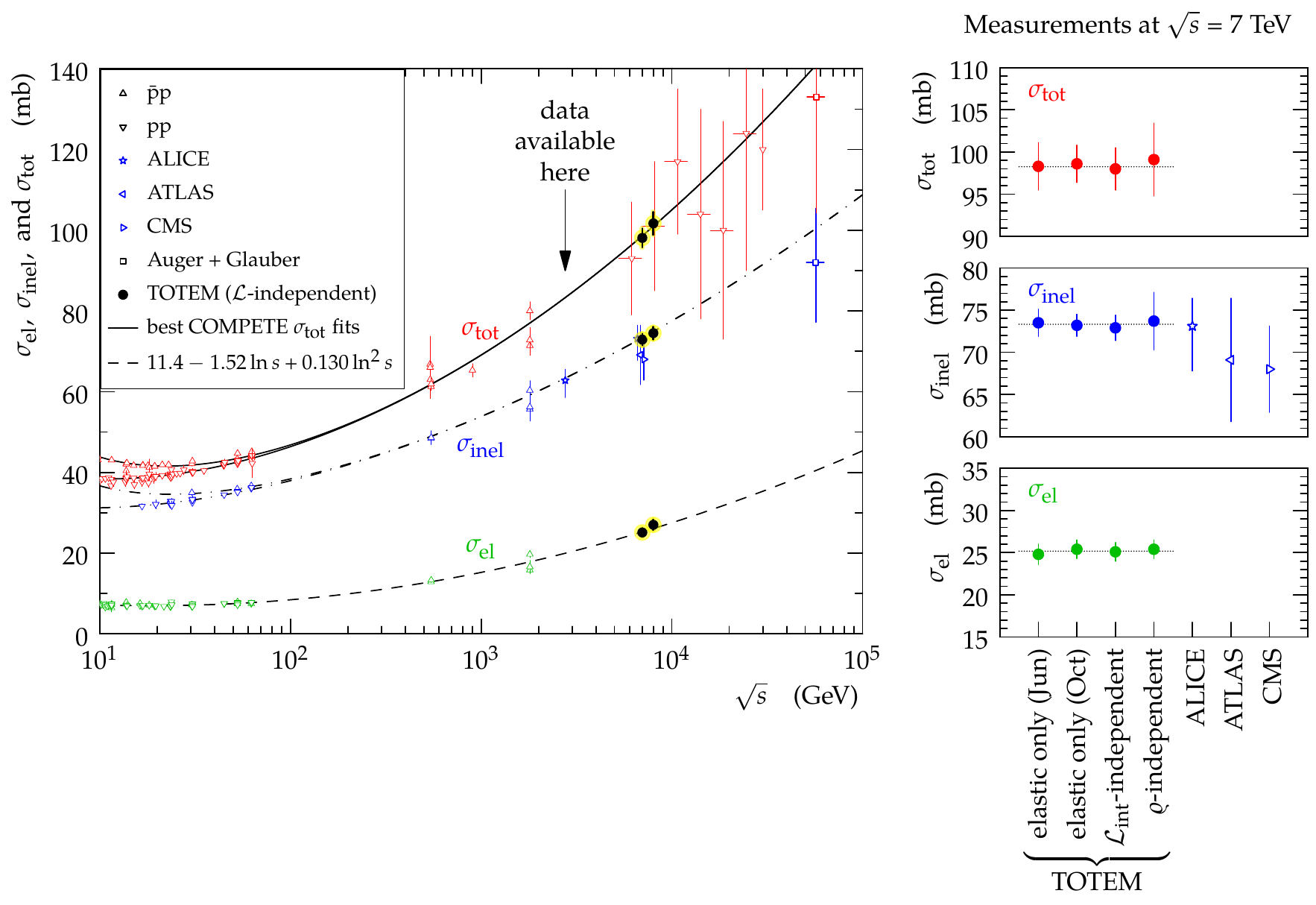}\pdfrefximage\pdflastximage}
\caption{%
{\em Left}: compilation \cite{si_tot_8,si_tot_7,pdg,alice_inel,atlas_inel_7,cms_inel_7,auger} of total, inelastic and elastic cross-sections plotted as a function of the centre-of-mass energy $\sqrt s$. The TOTEM measurements are highlighted in yellow. The continuous black lines (lower for $\rm pp$, upper for $\rm \bar pp$) represent the best fits of the total cross-section data by the COMPETE collaboration \cite{compete}. The dashed line results from a fit of the elastic scattering data. The dash-dotted curves correspond to the inelastic cross-section and were obtained as the difference between the continuous and dashed fits.\hfil\break
{\em Right}: detail of the measurements of total, inelastic and elastic cross-sections at $\sqrt s = 7\un{TeV}$. The circles represent the four TOTEM measurements, the other points show the measurements of other LHC collaborations.
}
\label{fig:sigmas}
\end{center}
\end{figure}

All the published total, elastic and inelastic cross-section results by TOTEM are summarised and put in context of earlier measurements in Figure \ref{fig:sigmas}.

\def\Name#1{#1,}
\def\Review#1#2#3#4{#1 {\bf#2} #4 (#3)}

\eject

\begin{footnotesize}

\end{footnotesize}
\end{document}